\documentclass[11pt]{article}
\usepackage{graphicx} 
\usepackage{geometry}
\usepackage[utf8]{inputenc}
\usepackage{amsmath}
\usepackage{amssymb}
\usepackage{float}
\usepackage[colorlinks=true, linkcolor=cyan, urlcolor=black, citecolor=cyan]{hyperref}
\usepackage{cite}
\usepackage{subcaption}
\usepackage[font=small,skip=0pt]{caption}
\usepackage{tikz}
\usetikzlibrary{positioning}
\usepackage{authblk}
\usepackage[font=small,skip=7pt]{caption}
\title{\textbf{Gravitational Lensing by a Dark Compact Object in Modified Gravity and Observational Constraints from Einstein Rings}}
\author[1]{Nagina Rehman \footnote{naginarehman.0123@gmail.com }} 
\author[2]{Zoe C S Chan \footnote{ bravychancs@gmail.com}}
\author[1]{Mubasher Jamil \footnote{ mjamil@sns.nust.edu.pk (corresponding author)}}
\author[3]{Mustapha Azreg-A\"{\i}nou\thanks{azreg@baskent.edu.tr}}

\affil[1]{\it{ School of Natural Sciences (SNS), National University of Sciences and Technology (NUST), H-12, 44000 Islamabad, Pakistan}}
\affil[2]{\it{Department of Physics, Xiamen University Malaysia, 43900 Sepang, Malaysia}}
\affil[3]{\it{Ba\c{s}kent University, Engineering Faculty, Ba\u{g}l{\i}ca Campus, 06790-Ankara, T\"{u}rkiye}}

\date{}

\begin{document}
\maketitle
\begin{abstract}
In this manuscript, we provide a comprehensive study of gravitational lensing by dark compact objects predicted by a Modified Gravity (MOG) based on the Scalar-Vector-Tensor action, and the aim is to analyze new insights into the nature of gravitational interactions. We compute weak and strong deflection angles for the specified static, spherically symmetric MOG spacetime. Additionally, we dedicate a section to explore observational implications in the weak field limit. By employing a supermassive galactic black hole as a gravitational lens, we compare various parameters in MOG with those of the Schwarzschild black hole as lens in strong-field scenarios. Specifically, we model the black holes M87${^*}$ and Sgr A${^*}$ as lenses within the MOG framework, calculating the corresponding lensing coefficients and distortion parameters in the weak field regime.
\end{abstract}
\newpage
\section{Introduction} 

Gravitational lensing proved to be a crucial observational tool for investigating the properties and distribution of baryonic and dark matter across a wide range of masses in galaxies, enabling significant advancements in our understanding of galactic dynamics and cosmic structures \cite{wambsganss1998gravitational}. Einstein introduced the notion of gravitational lensing in the early $20^{\text{th}}$ century as a consequence of his General Theory of Relativity (GR). He demonstrated that large celestial bodies could bend the path of light due to the curvature of spacetime \cite{gribbin2015einstein}. Eddington confirmed this prediction during his exploration of a solar eclipse in 1919 \cite{einstein1905does, bauer1920resume, earman1980relativity}. Later, Chwolson noted that a massive object could bend light from a more distant source and multiple images of that source would be formed \cite{shapiro1999century}.

The problems of dark matter and dark energy have produced a need for modified gravitational theories by suitably modifying GR. Moffat proposed the scalar-tensor-vector-gravity (STVG) theory, which involves restructuring the geometric part of GR. A heavy vector field and three scalar fields (mass, effective gravitational constant, and vector field coupling) express gravitational implications of spacetime in a MOG setup \cite{nozari2023accretion}. Much research has been carried out in the context of the modified gravity theory. Since then in 2015, researchers developed a modified gravity solution that avoids singularities and represents a dark compact object \cite{moffat2015modified}. Later, studies on the shadows of rotating and non-rotating black holes in modified gravity were done and found that these shadows depend on the mass and angular momentum \cite{moffat2015black}. Additionally, a rotating black hole solution in modified gravity was derived \cite{moffat2021modified}, which reduces to the familiar Schwarzschild and Kerr black holes when the parameter $\alpha$ is set to zero.

Gravitational lensing is an important astrophysical phenomenon for studying distant objects, quasars, and clusters of galaxies to estimate their mass and search exoplanets in various celestial systems \cite{paul2020strong,bozza2003quasiequatorial}. This concept has been widely used in the past few years to observe Einstein's rings, arcs, and the images of the background source. The study of dark matter in halos has been extensively investigated, particularly through strong lensing observations \cite{sarbu2001strong,chaturvedi2023exact}, and recent research on supermassive black holes in Bardeen gravity \cite{islam2024strong}. 
Other studies have explored the structure of dark matter, how gravitational lensing changes the procedures to measure the mass of galaxy clusters, and how dark energy might affect lensing \cite{Younas:2015sva,Azreg-Ainou:2017obt,vachher2024probing, meena2024gravitational, adil2024omnipotent}.

Weak gravitational lensing has been the focus of numerous studies in recent years \cite{pantig2020weak,arakida2012effect,ovgun2019weak}, based on the assumption that the deflection angle is small. These findings are in agreement with the predictions of GR. The weak gravitational lensing of dark compact objects within the framework of MOG remains an unexplored area of study. In 1937, Zwicky calculated the transverse mass and discussed the importance of observation of gravitational lensing \cite{sharma2014gravitational}. In 1963, Refsdal documented the properties of point mass gravitational lens and proposed that geometrical optics can be applied for gravitational lensing effects \cite{refsdal1992gravitational}. Walsh and Carswell first observed two identical quasar spectra, suspecting they were images of the same quasar, Q0957+561 \cite{walsh1979}. Later, Weymann confirmed that these spectra were gravitationally lensed images of a single quasar. This discovery, made in 1979, marked the identification of the first gravitational lens system \cite{stockton1980lens}. Following this breakthrough, the study of strong gravitational lensing experienced rapid growth, as multiple images of a source produced by lensing began offering valuable information about both the lens and the source \cite{zensus1995parsec,narlikar1983single}. A new Lens equation was also developed by Virbhadra and Ellis which is convenient for studying observable such as magnification in strong lensing \cite{virbhadra2000schwarzschild}. Later, Bozza \cite{bozza2002gravitational} developed the analytical method to obtain the expression of deflection angle in strong gravitational lensing, which was later refined by Tsukumoto \cite{tsukamoto2017deflection} and has been extended to find the analytical expression of deflection angles for several black holes \cite{kumaran2022deflection,tsukamoto2020nonlogarithmic}. Furthermore, the capture of the first horizon-scale image of the supermassive black hole in the M87 galaxy by the Event Horizon Telescope (EHT) has given a significant impetus to study gravitational lensing in the strong deflection limit (SDL) \cite{collaboration2019first,savolainen2019first,akiyama2019first}. Other theories have been also explored to study the effects of gravitational lensing and their results are consistent with GR  \cite{afrin2021parameter,rayimbaev2022shadow}. These researches motivate us to study the gravitational lensing by a dark compact object in MOG and to investigate observables in strong and weak gravity regimes by taking M87${^*}$ and Sgr A${^*}$ as reference lenses.

This paper is organized as follows: In Section \ref{2'}, we present the line element of a dark compact object in modified gravity. The calculation of the weak deflection angle of a dark compact object in MOG is carried out in Section \ref{3'}. Section \ref{4} is dedicated to the study of observable in weak filed limit with the graph of magnification and distortion parameter introduced by Virbhadra in \cite{virbhadra2022distortions}. The SDL of the spacetime is studied in Section \ref{5} and lensing coefficients $\overline{a}$ and $\overline{b}$ are plotted against $b$. In Section \ref{6}, the observables of SDL which include the Einstein's ring radius, deflection angles, the images and their separation for different values of free parameter $\alpha$ in MOG are examined alongside constraints on $\alpha$ against observations. We have taken M87${^*}$ and Sgr A${^*}$ as a lens and compared the results with GR. In Section \ref{7}, we concluded the whole paper. We write all equations in geometric units $G=c=1$ 
and only show second-order approximations to simplify the analytic calculations throughout the paper.

\section{Dark Compact Object in Modified Gravity}\label{2'}
The action for matter-free MOG gravitational theory, which includes nonlinear field equations for the gravitational spin-1 vector field \( \varphi^\mu \), is expressed as \cite{moffat2021regular}

\begin{equation}
    S_{\text{MOG}} = \frac{1}{16\pi G} \int d^4x \sqrt{-g} \left[ R - \mathcal{L}(B) \right],
\end{equation}
where \( \mathcal{L}(B) \) represents the Lagrangian density that accounts for the nonlinear contribution of
\begin{equation}
    B_{\mu\nu} = \partial_\mu \varphi_\nu - \partial_\nu \varphi_\mu,
\end{equation}
with $B=\frac{1}{4}B_{\mu\nu} B^{\mu\nu}$. In the absence of matter, the MOG field equations are given by:
\begin{equation}
    G_{\mu\nu} = -8\pi G T_{(\varphi)\mu\nu},
\end{equation}
\begin{equation}
    \nabla_\nu (B^{\mu\nu} L_B) = 0,
\end{equation}
\begin{equation}
    \nabla_\nu (^* B^{\mu\nu}) = 0,
\end{equation}
where \( ^*B_{\mu\nu} \) denotes the dual tensor of \( B_{\mu\nu} \). The gravitational energy-momentum tensor is expressed as \cite{moffat2021regular}
\begin{equation}
    T_{(\varphi)\mu\nu} = -\frac{1}{4\pi} \left[ g_{\mu\nu} \mathcal{L}(B) - B_{\mu \alpha} B _\nu \hspace{0.cm} ^\alpha L_B \right],
\end{equation}
where \( L_B = \frac{d \mathcal{L}(B)}{d B} \).
The gravitational constant is given by \( G = G_N(1+\alpha) \), or in natural units with \( G_N = 1 \), it simplifies to \( G = 1+\alpha \). The gravitational source charge for the spin-1 gravitational vector field \( \varphi_\mu \) is defined as:
\begin{equation}
    Q_g = \sqrt{\alpha} M,
\end{equation}
where \( M \) represents the mass of the compact object.  
From the field equations, the gravi-electric field is obtained as:
\begin{equation}
    E_g(r) = B_{01}(r) = -B_{10}(r).
\end{equation}
This results in the components of the energy-momentum tensor:
\begin{equation}
    T^0_{\varphi 0} = T^1_{\varphi 1} = -\frac{1}{4\pi} \left( \mathcal{L}(B) + E_g^2 L_B \right).
\end{equation}
Taking $\mathcal{L} =-B$ \cite{moffat2015modified}, the solution of the above field equations yields the line element of a static and spherically symmetric
spacetime in MOG as \cite{moffat2021regular}
\begin{equation}\label{1}
ds^2=A(r) dt^2-B(r) dr^2-C(r)d\Omega^2,
\end{equation}
here, $d\Omega^2=d\theta^2+\sin^2\theta d\phi^2$, and the functions $A(r)$, $B(r)$ and $C(r)$ are given as follows
\begin{equation}\label{2}
A(r)=f(r),\hspace{1cm}
B(r)=\frac{1}{f(r)},\hspace{1cm}
C(r)=r^2,
\end{equation}
where
\begin{equation}\label{2b}
f(r)=1-\frac{2\left(1+\alpha\right)Mr^2}{\left(r^2+\alpha(1+\alpha)M^2\right)^\frac{3}{2}}+\frac{\alpha(1+\alpha)M^2 r^2}{\left(r^2+\alpha(1+\alpha)M^2\right)^2}.
\end{equation}
In the limit $r \rightarrow \infty$, we get \cite{moffat2021regular}
\begin{equation}\label{3}
f(r)=1-\frac{2(1+\alpha)M}{r}+\frac{\alpha(1+\alpha)M^2}{r^2}+\mathcal{O}(M^3, \alpha^3).
\end{equation}
\(\alpha\) is defined as a dimensionless parameter \cite{moffat2021regular}. When we set \(\alpha = 0\) in (\ref{1}) and (\ref{2}), the Schwarzschild spacetime emerges from the line element which is associated with a static and spherically symmetric black hole. 
The parameter $\alpha$ has a critical value $\alpha_{crit}=0.674$. For $\alpha \leq \alpha_{crit}$, the MOG spacetime has two event horizons \cite{nozari2023accretion}
\begin{equation}\label{horizrad}
r_{\pm} = M \left(1 + \alpha \pm \sqrt{1 + \alpha} \right).
\end{equation}
When $\alpha > \alpha_{crit}$, it describes a horizonless and singularity-free MOG dark compact object. Notice from (\ref{horizrad}), the two event horizons merge when $\alpha=-1$, giving an extreme-like black hole. Then, $\alpha<-1$ describes a horizonless naked singularity akin to the Reissner-Nordstr\"{o}m \cite{reissner,nordstrom} and Kerr \cite{kerr} metric.
\section{Weak Deflection Angle of a Spacetime in MOG}\label{3'}

To calculate the weak deflection angle of static spherically symmetric spacetime in MOG, the Gibbons and Werner method will be adapted \cite{gibbons2008applications}. According to their approach, for asymptotically flat spacetime, the following formula is applicable to find the weak deflection angle 
\begin{equation}
   \alpha= -\int\int_{M} K d\sigma.
\end{equation}
Here, $d\sigma=\sqrt{\overline{g}_{rr} \overline{g}_{\phi\phi}}~dr d\phi$ is an areal element of the optical metric and $K$ is the Gaussian curvature of the 2-dimensional surface $M$ bounded by the light geodesic $r=r_\gamma(\phi)$, extending from the source to the observer, and by a circular path centered on the central object and extends to a large distance away from the central object (to ``spatial infinity''~\cite{gibbons2008applications}). Using the general notation of~\cite{arakida2018light}, the optical metric related to~\eqref{1}, is given by
\begin{equation}\label{om}
dt^2=\overline{g}_{rr} dr^2+\overline{g}_{\phi\phi} d\phi^2=\frac{B(r)}{A(r)}dr^2+\frac{C(r)}{A(r)}d\phi^2,
\end{equation}
where $B/A=1/(f(r))^2$ and $C/A=r^2/f(r)$. Further, $K$ is defined in terms of the optical metric by \cite{Maryam:2024wwd}
\begin{equation}\label{K}
K=-\frac{1}{\sqrt{\overline{g}_{rr} \overline{g}_{\phi\phi}}}~\frac{d}{dr}\Big(\frac{1}{\sqrt{\overline{g}_{rr}}}~\frac{d\sqrt{\overline{g}_{\phi\phi}}}{dr}\Big),
\end{equation}
yielding
\begin{equation}\label{kk1}
\hat{\alpha}=-\int_0^\pi \int_{r_\gamma(\phi)}^\infty K d\sigma=\int_0^\pi \int_{r_\gamma(\phi)}^\infty \frac{d}{dr}\Big(\frac{1}{\sqrt{\overline{g}_{rr}}}~\frac{d\sqrt{\overline{g}_{\phi\phi}}}{dr}\Big)drd\phi\,.
\end{equation}
Thus,
\begin{equation}\label{int1}
\hat{\alpha}=\int_0^\pi \bigg[\lim_{r\to\infty}\Big(\frac{1}{\sqrt{\overline{g}_{rr}}}~\frac{d\sqrt{\overline{g}_{\phi\phi}}}{dr}\Big)-\Big(\frac{1}{\sqrt{\overline{g}_{rr}}}~\frac{d\sqrt{\overline{g}_{\phi\phi}}}{dr}\Big)\Big|_{r=r_\gamma(\phi)}\bigg]d\phi\,.
\end{equation}
The integrand
\begin{equation}\label{int2}
\frac{1}{\sqrt{\overline{g}_{rr}}}~\frac{d\sqrt{\overline{g}_{\phi\phi}}}{dr}=\frac{\sqrt{A}}{\sqrt{B}}\frac{d}{dr}\Big(\frac{\sqrt{C}}{\sqrt{A}}\Big)=\frac{1}{2}\frac{\sqrt{C}}{\sqrt{B}}\frac{d}{dr}\Big(\ln\frac{C}{A}\Big)\,,	
\end{equation}
approaches unity in the limit $r\to\infty$ if the metric~\eqref{1} is asymptotically flat. In this case, the integrand admits a series expansion in powers of $1/r$ of the form:
\begin{equation}\label{int3}
\frac{1}{\sqrt{\overline{g}_{rr}}}~\frac{d\sqrt{\overline{g}_{\phi\phi}}}{dr}=1-\frac{a_\ell}{r^\ell}-\frac{a_m}{r^m}-\frac{a_n}{r^n}-\frac{a_p}{r^p}-\cdots,
\end{equation} 
where $\ell <m<n<p<\cdots$ are positive numbers, which are integers for most known solutions ($\ell =1,\, m=2,\,n=3,\,p=4,\,\cdots$), and $a_\ell$, $a_m$, $a_n,\,a_p,\,\cdots$ are real numbers. The expression of $\hat{\alpha}$ takes the form
\begin{equation}\label{int4}
\hat{\alpha}=\int_0^\pi \bigg[\frac{a_\ell}{r_{\gamma}(\phi)^\ell}+\frac{a_m}{r_{\gamma}(\phi)^m}+\frac{a_n}{r_{\gamma}(\phi)^n}+\frac{a_p}{r_{\gamma}(\phi)^p}+\cdots\bigg]d\phi\,.	
\end{equation}
The light geodesic $r=r_\gamma(\phi)$, which satisfies the differential equation,
\begin{equation}\label{de}
\Big(\frac{dr}{d\phi}\Big)^2=\frac{\overline{g}_{\phi\phi}}{\overline{g}_{rr}}\Big(\frac{\overline{g}_{\phi\phi}}{b^2}-1\Big)=\frac{C}{B}\Big(\frac{C}{b^2A}-1\Big)\,,
\end{equation}
admits a similar expansion in powers of the impact parameter $b$ of the form
\begin{equation}\label{int5}
\frac{1}{r_\gamma}=\frac{g_s(\phi)}{b^s}+\frac{g_u(\phi)}{b^u}+\frac{g_v(\phi)}{b^v}+\cdots,
\end{equation} 
where $s <u<v<\cdots$ are positive numbers, which are integers for most known solutions ($s =1,\, u=2,\,v=3,\,\cdots$), and $g_s(\phi)$, $g_s(\phi)$, $g_s(\phi),\,\cdots$ are functions of $\phi$. Upon substituting~\eqref{int5}  into~\eqref{int4}, we obtain $\hat{\alpha}$ up to desired order in $1/b$.

\begin{figure}[ht]
    \centering
    \includegraphics[width=0.6\textwidth]{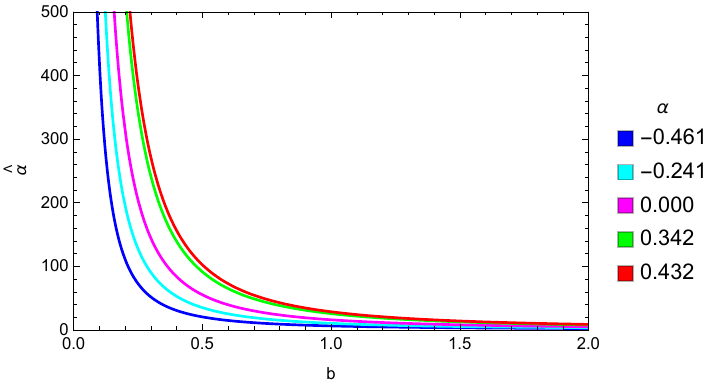}
    \caption{Graph of weak deflection angle for different values of $\alpha$ is plotted here. The magenta curve represents the Schwarzschild black hole case for $\alpha=0$. }
    \label{10}
\end{figure}

For MOG, using the general expression for $f(r)$~\eqref{2b}, we obtain
\begin{equation}\label{mog1}
\frac{a_\ell}{r_{\gamma}(\phi)^\ell}+\frac{a_m}{r_{\gamma}(\phi)^m}+\frac{a_n}{r_{\gamma}(\phi)^n}+\cdots = \frac{2M(1+\alpha)}{r_{\gamma}(\phi)}+\frac{3}{2}~\frac{M^2(1+\alpha)}{r_{\gamma}(\phi)^2}+\frac{2M^3(1+\alpha)^2(1-3\alpha)}{r_{\gamma}(\phi)^3}+\cdots,\,
\end{equation}
and
\begin{equation}\label{kk2}
	\frac{1}{r_\gamma}=\frac{g_s(\phi)}{b^s}+\frac{g_u(\phi)}{b^u}+\cdots = \frac{\sin\phi}{b}+\frac{M(1+\alpha)(1-\cos\phi)^2}{2b^2}+\cdots\,,
\end{equation}
where we fixed the integration constant taking $r_\gamma(0)=\infty$. Up to $\alpha^2/b^2$, we obtain
\begin{equation}\label{44}
{\hat{\alpha}}=\frac{4M}{b}+\frac{15\pi M^2}{4b^2}+\left(\frac{4M}{b}+\frac{27\pi M^2}{4b^2}\right)\alpha+\frac{3\pi M^2\alpha^2}{b^2}+\cdots
\end{equation}
The weak deflection angle for the Schwarzschild black hole~\cite{pantig2020weak} is recovered upon substituting $\alpha =0$ in this last expression.
For any asymptotically flat, and spherically symmetric, metric~\eqref{1}, the general procedure, for determining the deflection angle, consists in expanding~\eqref{int2}
\begin{equation*}
\frac{1}{2}\frac{\sqrt{C}}{\sqrt{B}}\frac{d}{dr}\Big(\ln\frac{C}{A}\Big)
\end{equation*}
in powers of $1/r$ and expanding the solution to the differential equation~\eqref{de} 
\begin{equation*}
\Big(\frac{dr}{d\phi}\Big)^2=\frac{C}{B}\Big(\frac{C}{b^2A}-1\Big),
\end{equation*}
in powers of $1/b$ with coefficients depending on $\phi$, substituting the latter expansion into the former one and integrating over $\phi$ from $0$ to $\pi$.
The weak deflection angle of an incoming photon\footnote{An ``incoming photon" refers to a photon that is approaching a given point or observer in space, typically in the context of light propagation and detection.} from the asymptotic region in MOG is plotted against $b$ as shown in Figure \ref{10}. The graph is drawn for different values of $\alpha$. As $\alpha$ increases, the bending angle $\hat{\alpha}$ increases accordingly.

\section{Observables in Weak Field Limit}\label{4}
\
The observables in MOG are discussed by taking the mass and distance of M87{$^*$}. In Modified Gravity (MOG), studying observables in the weak field limit is crucial for testing how the theory deviates from GR. To better understand the shapes of images formed in modified gravity, we examine the variations in tangential and radial magnifications, as well as the total magnifications of the primary, secondary, and first-order relativistic images (on both sides of the optic axis) against $\beta$, $\mathcal{D}$, and $\frac{M}{D_{ol}}$. $\mathcal{D}$ represents the ratio of lens-source to the observer-source distance in the lensing configuration, and $D_{OL}$ is the distance between the observer and the lens.

\begin{figure}[ht]
    \centering
    \begin{subfigure}{0.3\textwidth}
        \centering
        \includegraphics[width=\textwidth]{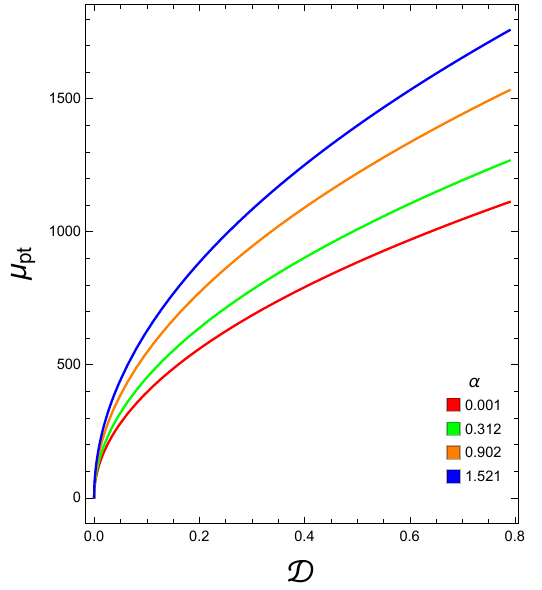}
        \caption{}
        \label{fig.1}
    \end{subfigure}
    \hfill
    \begin{subfigure}{0.31\textwidth}
        \centering
        \includegraphics[width=\textwidth]{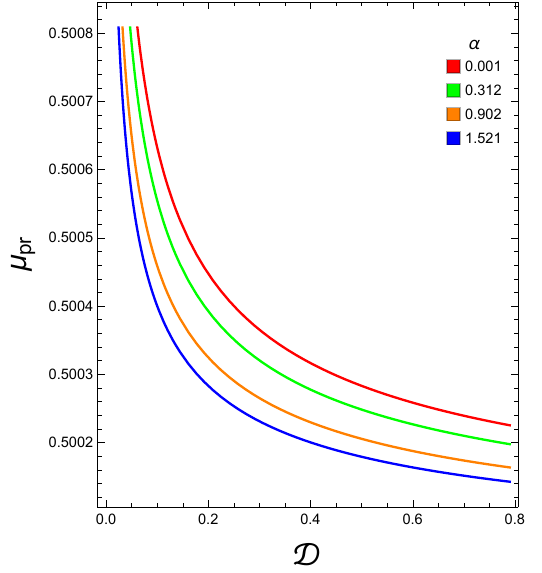}
        \caption{}
        \label{fig.2}
    \end{subfigure}
    \hfill
    \begin{subfigure}{0.3\textwidth}
        \centering
        \includegraphics[width=\textwidth]{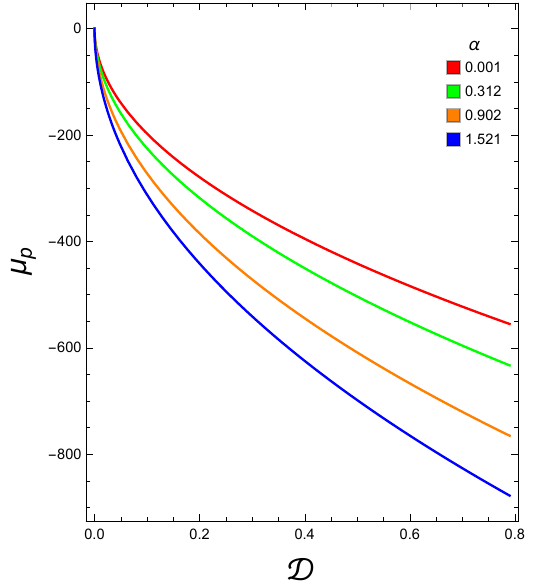}
        \caption{}
        \label{fig.3}
    \end{subfigure}
    \hfill
    \begin{subfigure}[b]{0.3\textwidth}
        \centering
        \includegraphics[width=\textwidth]{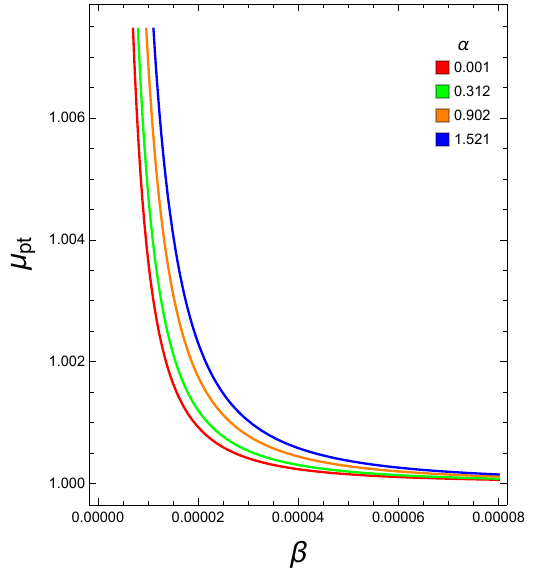}
        \caption{}
        \label{fig:fig15}
    \end{subfigure}
    \hfill
    \begin{subfigure}[b]{0.3\textwidth}
        \centering
        \includegraphics[width=\textwidth]{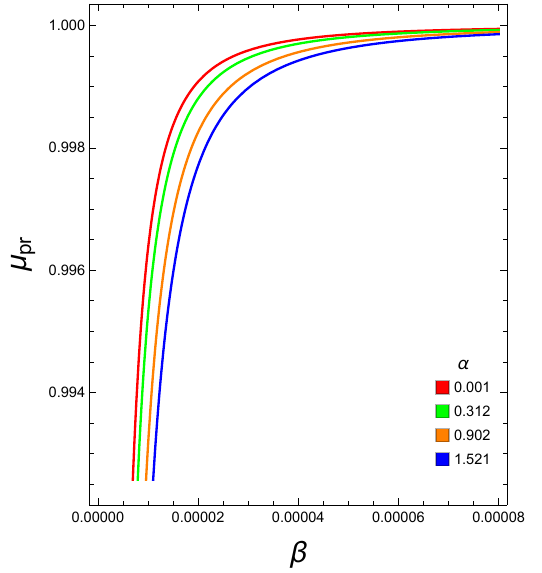 }
        \caption{}
        \label{fig:fig16}
    \end{subfigure}
    \hfill
    \begin{subfigure}[b]{0.31\textwidth}
        \centering
        \includegraphics[width=\textwidth]{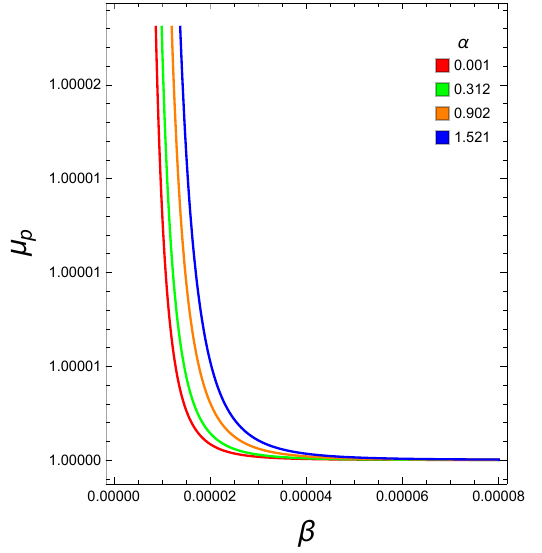}
        \caption{}
        \label{fig:fig17}
    \end{subfigure}
    \hfill
    \begin{subfigure}[b]{0.3\textwidth}
        \centering
        \includegraphics[width=\textwidth]{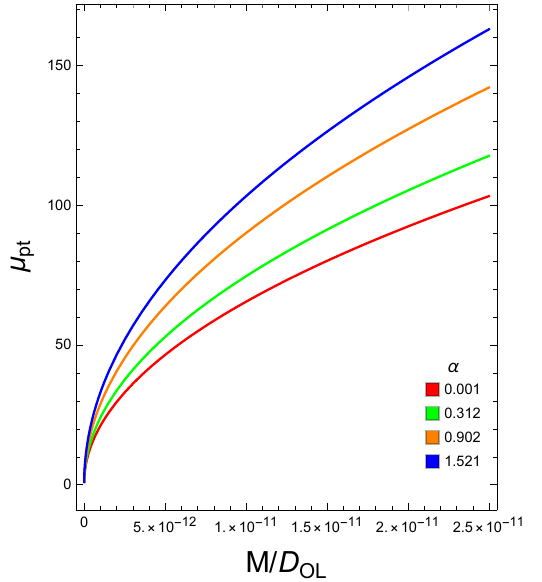}
        \caption{}
        \label{fig:fig18}
    \end{subfigure}
    \hfill
    \begin{subfigure}[b]{0.31\textwidth}
        \centering
        \includegraphics[width=\textwidth]{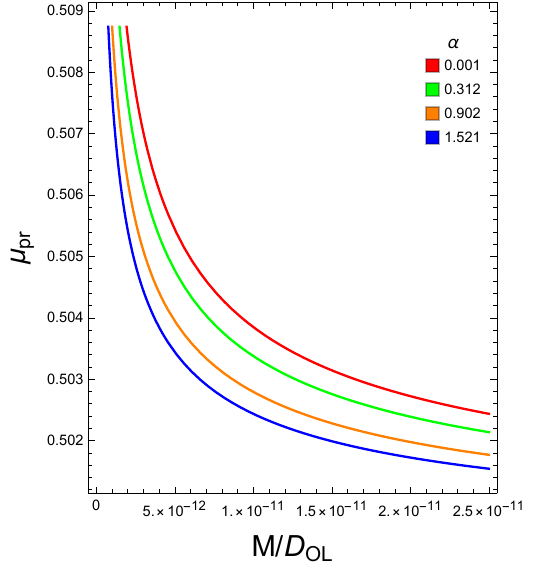}
        \caption{}
        \label{fig:fig19}
    \end{subfigure}
    \hfill
    \begin{subfigure}[b]{0.3\textwidth}
        \centering
        \includegraphics[width=\textwidth]{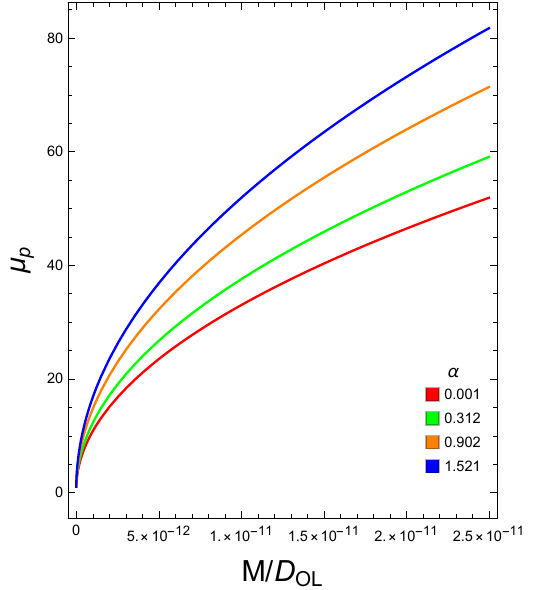}
        \caption{}
        \label{fig:fig20}
    \end{subfigure}
    \caption{The tangential, radial, and total magnifications are plotted against different parameters for primary images. We have assigned different values to the free parameter $\alpha$. The graphs of tangential magnification $\mu_{pt}$, radial magnification $\mu_{pr}$, and total magnification of primary images $\mu_p$ are drawn for distinct values of $\mathcal{D}$, $\beta$ and $M/D_{OL}$.}
    \label{fig:all4}
\end{figure}

\subsection{Lens Equation, Magnification and Distortion Parameter}
\

To explore the phenomenon of gravitational lensing in both weak and strong gravitational fields, a lens equation was proposed in  \cite{virbhadra2000schwarzschild} as
\begin{equation}\label{g}
    \tan\beta=\tan\theta-\alpha,
\end{equation}
where
\

\begin{center}
  $\alpha=\frac{D_{LS}}{D_{OS}}\tan(\hat{\alpha}-\theta)$.
\end{center}
The angular positions of the image and the source are represented by the symbols $\theta$ and $\beta$, respectively. Here, $D_{OS}$ refers to the distance between the observer and the source, while $D_{LS}$ is the distance between the lensing object and the source. The Einstein deflection angle of the light ray is denoted by $\hat{\alpha}$. The angular position of the Einstein ring is obtained by setting $\beta = 0$ in \eqref{g}. The impact parameter can also be expressed as a function of $\theta$ given by
\begin{equation}
    b=D_{OL} \sin \theta.
\end{equation}
$D_{OL}$ is the observer-lens distance.
The radial magnification  $\mu_r$, the tangential magnification $\mu_t$, and total magnification $\mu$ of image can be expressed as 

\begin{equation}\label{h}
     \mu_r=\left(\frac{d\beta}{d\theta}\right)^{-1},\hspace{1cm} \mu_t=\left(\frac{\sin \beta}{\sin \theta}\right)^{-1} ,\hspace{1cm}\mu=\mu_t \mu_ r.
\end{equation}
The distortion parameter is defined in \cite{virbhadra2022distortions} as
\begin{equation}\label{i}
    \Delta=\frac{\mu_t}{\mu_r}.
\end{equation}
Also, a logarithmic distortion parameter of the image is introduced in \cite{virbhadra2022distortions} which has a mathematical expression
\begin{equation}\label{j}
  \delta=\log_{10} \bigg|\frac{\mu_t}{\mu_r}\bigg|.
\end{equation}
We will consider the simplest gravitational lensing i.e., the weak field lensing in modified gravity. \eqref{g} reduces to the well-known form \cite{2006glsw.conf.....M}
\begin{equation}\label{k}
    \beta=\theta- \hat{\alpha} \mathcal{D}.
\end{equation}
where the bending angle  for the spacetime in MOG becomes
\begin{equation*}
    \hat{\alpha}=\frac{4M}{b}(1+\alpha)+O(\alpha^2, M^2).
\end{equation*}
Solving \eqref{k} gives the angular position of primary and secondary images as
\begin{equation}\label{l}
\begin{split}
     \theta_p &=\frac{1}{2}\left(\beta+\sqrt{\beta^2+4 \theta_E^2}\right), \hspace{1cm} \\
    \theta_s &=\frac{1}{2}\left(\beta-\sqrt{\beta^2+4 \theta_E^2}\right),
\end{split}
\end{equation}
where $\theta_E$ is the angular radius of the Einstein ring. Its mathematical expression in MOG is given by
\begin{equation}\label{mogring}
    \theta_E=\sqrt{\frac{4\mathcal{D}M}{D_{OL}}\left(1+\alpha\right)} .
\end{equation}
Using \eqref{h} and \eqref{l} in \eqref{j}, the distortion of the primary and secondary images is obtained as 
\begin{equation}\label{o}
    \Delta_p=-\Delta_s=\bigg|\frac{\sqrt{\beta^2+16\mathcal{D} \frac{M}{D_{OL}}}}{\beta}\bigg|.
\end{equation}    
The subscripts $p$ and $s$ are used for the primary and secondary relativistic images respectively. The Virbhadra's hypothesis suggests that, in weak gravitational lensing, distortions in image positions cancel each other out, leading to zero net distortion \cite{virbhadra2022distortions}. Consequently,
\begin{equation}
    \Delta_{sum}=0.
\end{equation}
Using \eqref{j} and \eqref{l}, the magnification and distortion parameters for primary and secondary pictures formed in MOG are plotted for different values of $\alpha$ shown in Figures \ref{fig:all4}, \ref{fig:all2}, and \ref{fig:all3}. 
\subsection{Computations and Results}

\ 

To gain a deeper understanding of relativistic images formed through gravitational lensing in MOG, we have analyzed the tangential, radial, and total magnifications for both primary and secondary images. A distortion parameter is also plotted against $D$, $\beta$, and $M/D_{OL}$ to study the distortion of primary and secondary relativistic images. We have modeled M87${^*}$ (mass $M=6.5 \times 10^9$ solar mass and $D_{OL}=16.8$ M$pc$) as lens in MOG. The value of angular source position $\beta=1 \text{m}as$ is used in the numerical analysis.

\ 

\begin{figure}[h]
    \centering
    \begin{subfigure}{0.3\textwidth}
        \centering
        \includegraphics[width=\textwidth]{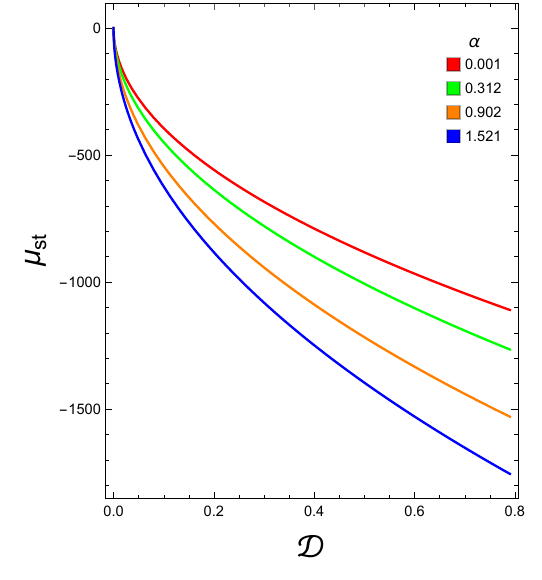}
        \caption{}
        \label{fig.4}
    \end{subfigure}
    \hfill
    \begin{subfigure}{0.31\textwidth}
        \centering
        \includegraphics[width=\textwidth]{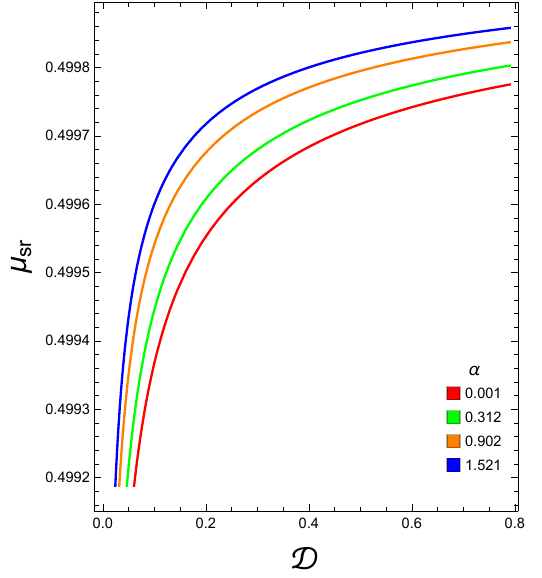}
        \caption{}
        \label{fig.5}
    \end{subfigure}
    \hfill
    \begin{subfigure}{0.3\textwidth}
        \centering
        \includegraphics[width=\textwidth]{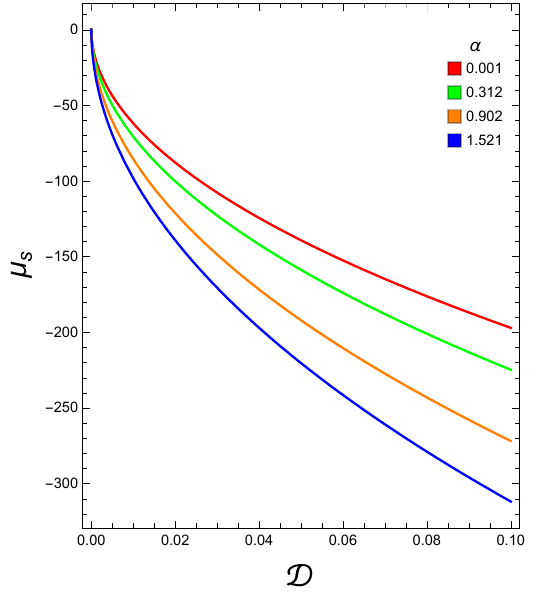}
        \caption{}
        \label{fig.6}
    \end{subfigure}
    \hfill
    \begin{subfigure}[b]{0.3\textwidth}
        \centering
        \includegraphics[width=\textwidth]{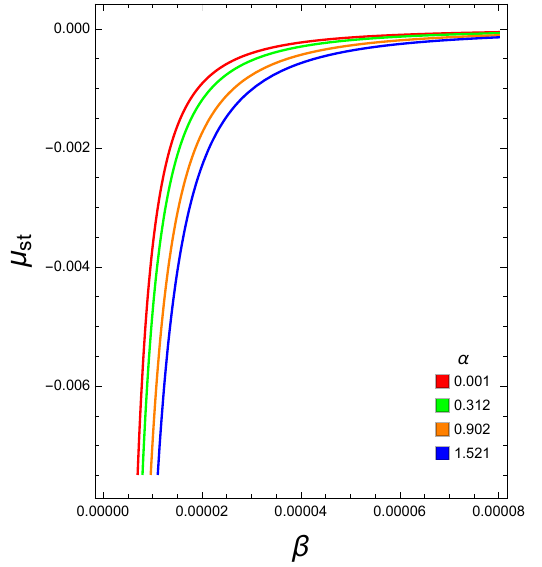}
        \caption{}
        \label{fig:fig7}
    \end{subfigure}
    \hfill
    \begin{subfigure}[b]{0.3\textwidth}
        \centering
        \includegraphics[width=\textwidth]{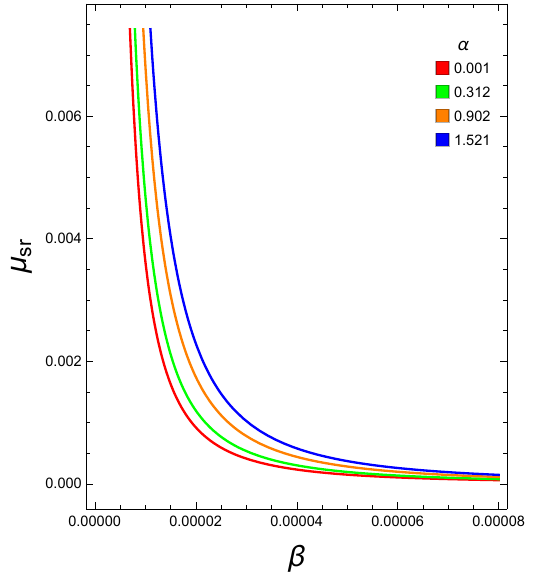}
        \caption{}
        \label{fig:fig8}
    \end{subfigure}
    \hfill
    \begin{subfigure}[b]{0.31\textwidth}
        \centering
        \includegraphics[width=\textwidth]{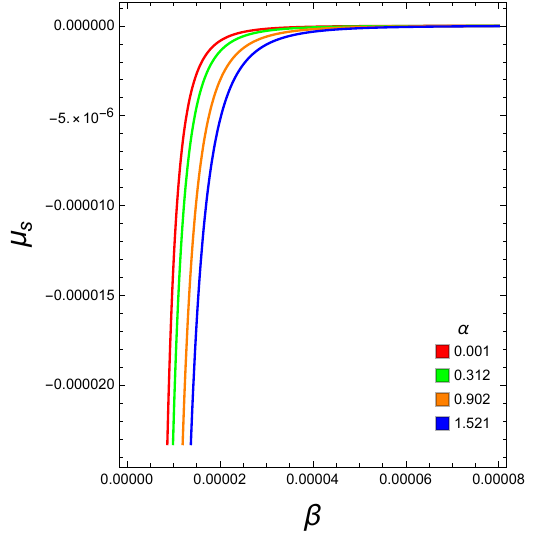}
        \caption{}
        \label{fig:fig9}
    \end{subfigure}
    \hfill
    \begin{subfigure}[b]{0.3\textwidth}
        \centering
        \includegraphics[width=\textwidth]{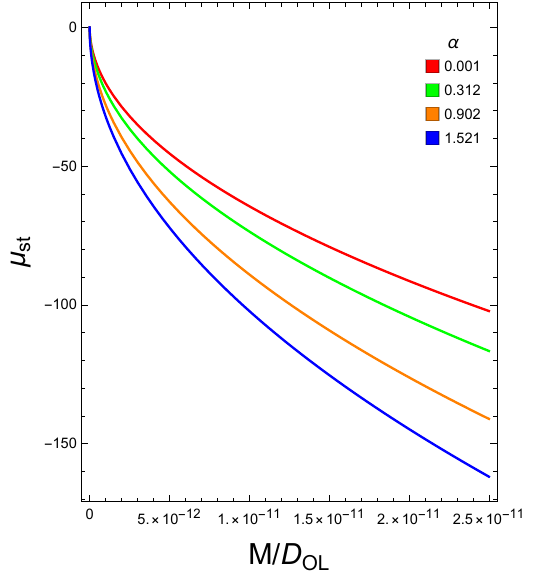}
        \caption{}
        \label{fig:fig10}
    \end{subfigure}
    \hfill
    \begin{subfigure}[b]{0.31\textwidth}
        \centering
        \includegraphics[width=\textwidth]{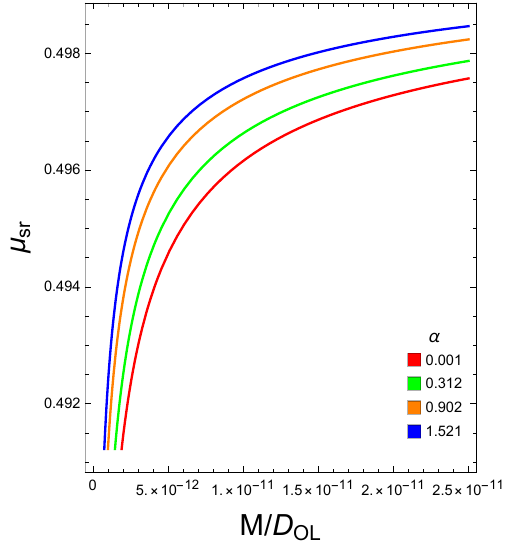}
        \caption{}
        \label{fig:fig11}
    \end{subfigure}
    \hfill
    \begin{subfigure}[b]{0.3\textwidth}
        \centering
        \includegraphics[width=\textwidth]{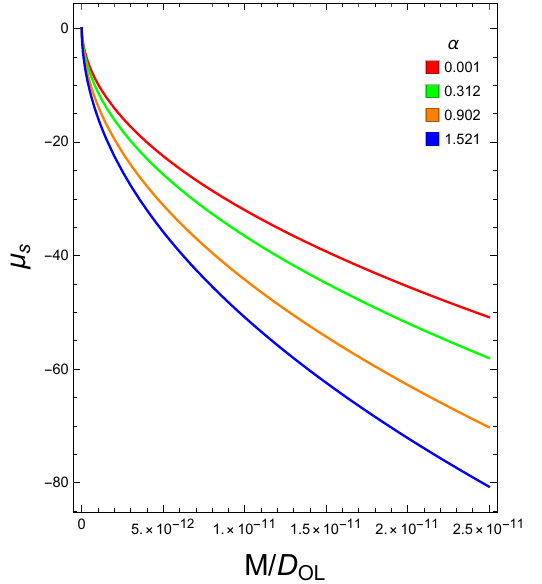}
        \caption{}
        \label{fig:fig12}
    \end{subfigure}
    \caption{For the secondary images, the tangential, radial, and total magnifications are plotted as functions of several parameters. Different values of $\alpha$ are used to examine the properties of the secondary images. The plots are produced for varying values of $\mathcal{D}$, $\beta$, and the ratio $M/D_{OL}$.}
    \label{fig:all2}
\end{figure}

\begin{figure}[h]
    \centering
    \begin{subfigure}{0.3\textwidth}
        \centering
        \includegraphics[width=\textwidth]{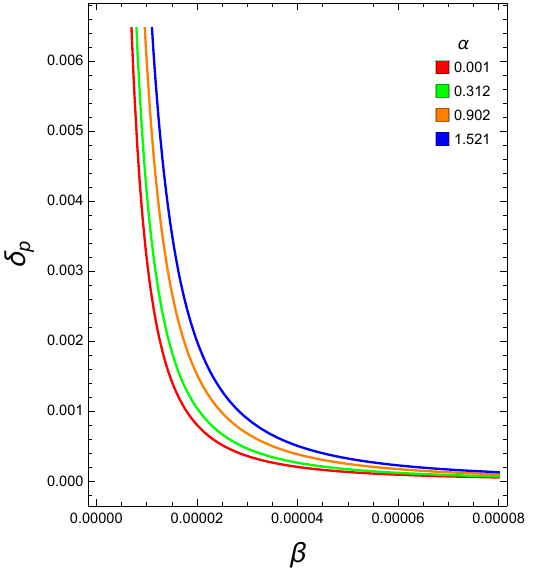}
        \caption{}
    \end{subfigure}
    \hfill
    \begin{subfigure}{0.29\textwidth}
        \centering
        \includegraphics[width=\textwidth]{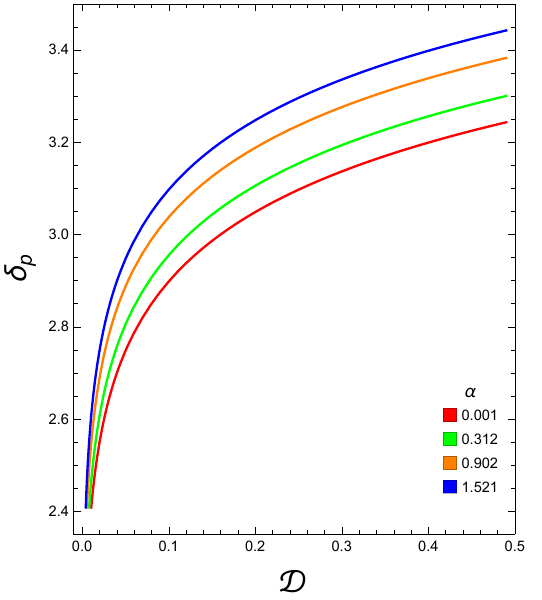}
        \caption{}
    \end{subfigure}
    \hfill
    \begin{subfigure}{0.3\textwidth}
        \centering
        \includegraphics[width=\textwidth]{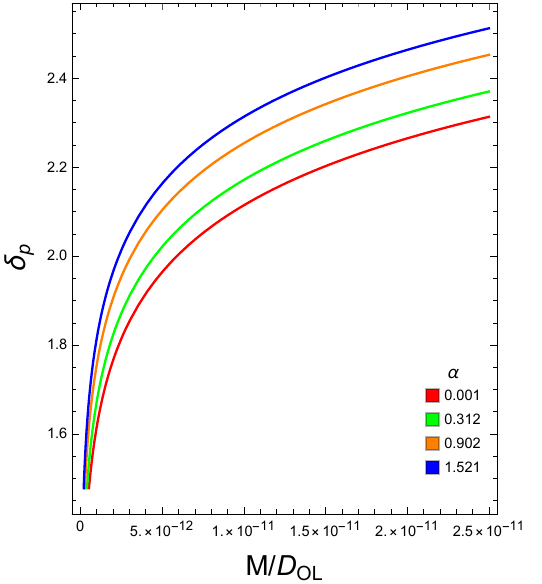}
        \caption{}
    \end{subfigure}
    \hfill
    \begin{subfigure}[b]{0.3\textwidth}
        \centering
        \includegraphics[width=\textwidth]{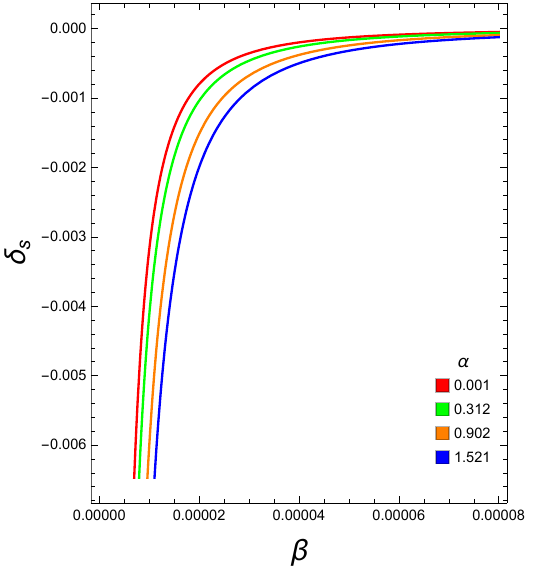}
        \caption{}
        \label{fig14}
    \end{subfigure}
    \hfill
    \begin{subfigure}[b]{0.3\textwidth}
        \centering
        \includegraphics[width=\textwidth]{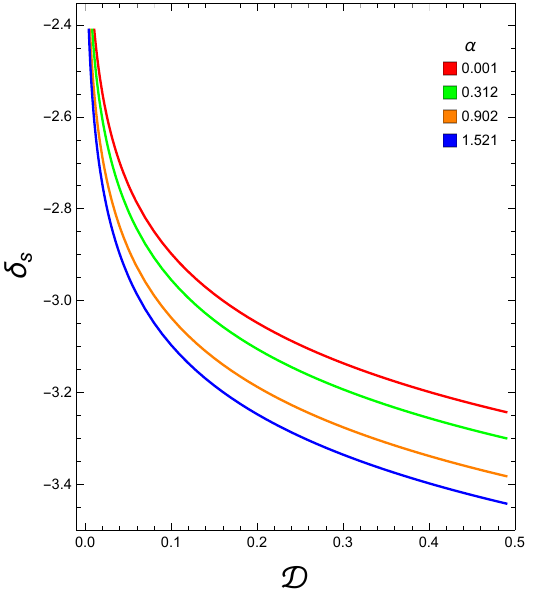 }
        \caption{}
        \label{fig:15}
    \end{subfigure}
    \hfill
    \begin{subfigure}[b]{0.3\textwidth}
        \centering
        \includegraphics[width=\textwidth]{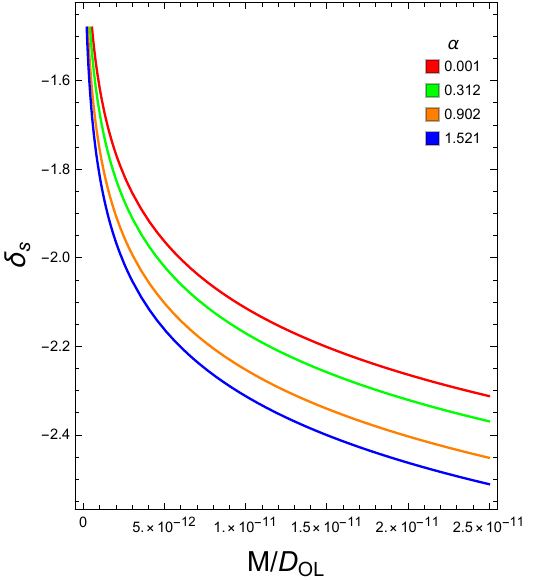}
        \caption{}
        \label{fig:16}
    \end{subfigure}
    
    \caption{The Distortion parameter is plotted for different values of $\mathcal{D}$, $\beta$, and $M/D_{OL}$. In the first row, the graphs of primary images are shown for different values of $\alpha$. The second row contains the graph of secondary images plotted against the parameters mentioned above.}
    \label{fig:all3}
\end{figure}

Figure \ref{fig:all4} illustrates the graphs related to tangential, radial, and total magnification for primary images. With an increase in $D$ and $\alpha$, $\mu_{pt}$ grows significantly. $\mu_{pr}$ and $\mu_p$ exhibit a decreasing pattern when $\alpha$ and $D$ increase. When graphs are plotted against $\beta$, $\mu_{pt}$ decreases with an increasing $\beta$ but increases with $\alpha$.  Conversely, $\mu_{pr}$ increases with an increasing $\beta$ and decreases with $\alpha$. The total magnification $\mu_p$ follows the behavior of $\mu_{pt}$. Further, when plots are generated against $M/D_{OL}$, the total magnification and tangential magnification of primary images increases with increasing $\alpha$ and $M/D_{OL}$. But, $\mu_{pr}$ shows the declining behavior and decreases with increasing $\alpha$ and $M/D_{OL}$. The secondary relativistic images in Figure \ref{fig:all2} show the same results as for primary relativistic images. The logarithmic distortion of primary and secondary images is plotted against different values of $D$, $\beta$, and $M/D_{OL}$ in Figure \ref{fig:all3}. Opposite behavior for the primary and secondary relativistic images can be observed through their plots.

\ 

\section{Deflection Angle in the Strong Field Limit}\label{5}
The strong deflection angle of dark compact objects in MOG can be calculated using the improved approach of Tsukamoto \cite{tsukamoto2017deflection}. The given spacetime is static, spherically symmetric, and asymptotically flat as it satisfies an asymptotically flat condition. Using \eqref{2}, we obtain
\begin{equation}
    \lim_{r\to\infty} A(r)=1,
\end{equation}
\begin{equation}
    \lim_{r\to\infty} B(r)=1,
\end{equation}
\begin{equation}
    \lim_{r\to\infty} \frac{C(r)}{r^2}=1.
\end{equation}

The expressions of the radius of the photon sphere $r_m$, impact parameter $b$, and critical impact parameter $b_c$ are calculated by following the procedure defined in \cite{tsukamoto2017deflection}. First of all, the radius of the photon sphere $r_m$ can be obtained by assuming that there exists at least one positive solution of $\text{D}(r)=0$ \footnote{$\text{D}(r)$ is derived by differentiating the effective potential with respect to $r$.}, where
\begin{equation}
    \text{D}(r)=\frac{C'}{C}-\frac{A'}{A}.
\end{equation}
By using the required values from \eqref{2}, we get
\begin{equation}
    r_m=\frac{M}{2}\left(3+3\alpha+\sqrt{9+10 \alpha+ \alpha^2}\right),
\end{equation}
 which is the radius of the photon sphere in MOG. 
Secondly, the perpendicular distance between the photon coming from the source and the center of a lensing body called the impact parameter is defined by 
\begin{equation}
b=\sqrt{\frac{C_t}{A_t}}.
\end{equation}
Here and hereafter the subscript $t$ denotes the quantities at turning point where $r=r_t$. By definition, $r_t$ is the closest approach distance of the light ray near the lensing object.
\clearpage

\begin{figure}[h]
    \centering
    \includegraphics[width=0.5\textwidth]{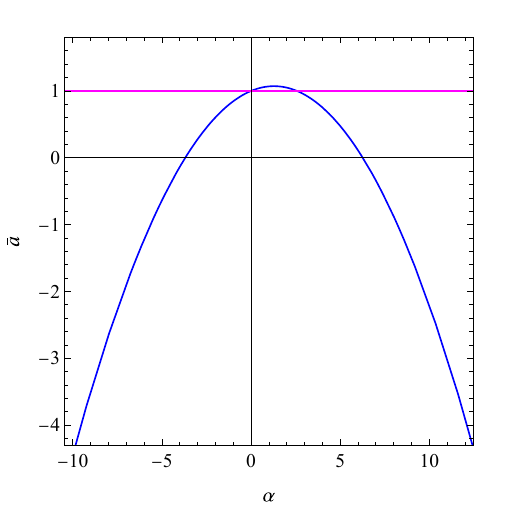}\hfill
    \includegraphics[width=0.5\textwidth]{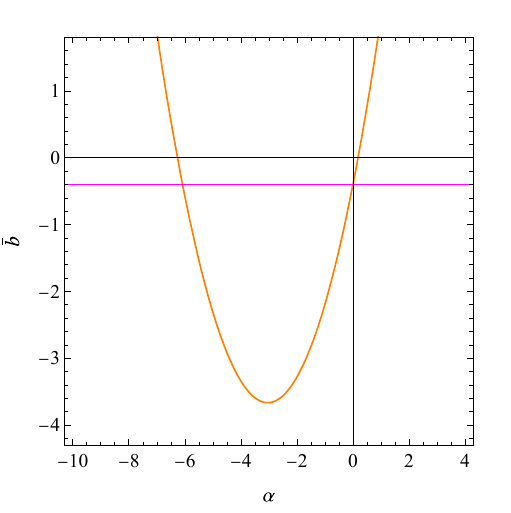}
    \caption{$\overline{a}$ and $\overline{b}$ is plotted against the different values of $\alpha$.
    Two $\alpha$ values correspond to a single $\overline{a}$ or $\overline{b}$ value except at 
    the turning point of each graphs; $\alpha=\frac{9}{7}$ for $\overline{a}$ and $\alpha=-3.04245$ for $\overline{b}$. Magenta straight lines represent the $\overline{a}$ and $\overline{b}$ values in the Schwarzschild case.}
    \label{fig:2}
\end{figure}

Critical impact parameter \footnote{The critical impact parameter is the minimum distance from which a photon can approach a massive object and escape to infinity.} has the following mathematical form \cite{bozza2010gravitational} 
\begin{equation}
b_c=\lim_{r_t\to\ r_m}\sqrt{\frac{C_t}{A_t}}.
\end{equation}
After using the required values in $b_c$ from \eqref{2}, we obtain
\begin{equation}\label{bc}
    b_c=3\sqrt{3}M+\frac{5}{2}\sqrt{3}M \alpha -\frac{7}{24\sqrt{3}}M\alpha^2+\mathcal{O}(M^3, \alpha^3).
\end{equation}
The light ray coming from the asymptotic region will deflect only if the impact parameter \( b \) is greater than the critical value \( b_c \), i.e., \(b > b_c.\) If \( b \leq b_c \), the light ray may either orbit the massive object or be absorbed by it, depending on its impact parameter.
Deflection angle is given by the expression \cite{tsukamoto2017deflection}
\begin{equation}\label{z2}
    \hat{\alpha}=\overline{a} \log\left(\frac{b}{b_c}-1\right)+\overline{b}+\mathcal{O}\left((b-b_c)\log(b-b_c)\right),
\end{equation}
where $\overline a$ and $\overline b$ are defined as 
\begin{equation}\label{v}
    \overline a=\sqrt{\frac{2B_m A_m}{C_m''A_m-A_m''C_m}},
\end{equation}
and
\begin{equation}\label{w}
    \overline b=\overline a \log\left[r_m^2\left(\frac{C_m''}{C_m}\frac{A_m''}{A_m}\right)\right]+I_R(r_m)-\pi.
\end{equation}
Substituting the values of $A(r)$, $B(r)$, and $C(r)$ from \eqref{2} into \eqref{v}, we get 
\begin{equation}\label{y}
    \overline a =1+\frac{\alpha}{9}- \frac{7 \alpha^2}{162}+\mathcal{O}(M^3, \alpha^3).
\end{equation}
Using equation \eqref{w}, the value of $\overline b$ is as follows
\begin{equation}\label{z}
\overline{b}= \log(6) + \frac{1}{9} \alpha \big(-1 + \log(6)\big) - \frac{1}{162} \alpha^2 \big(-5 + 7 \log(6)\big) + I_R(r_m) - \pi.
\end{equation}
\begin{figure}[h]
    \centering
    \includegraphics[width=0.7\textwidth]{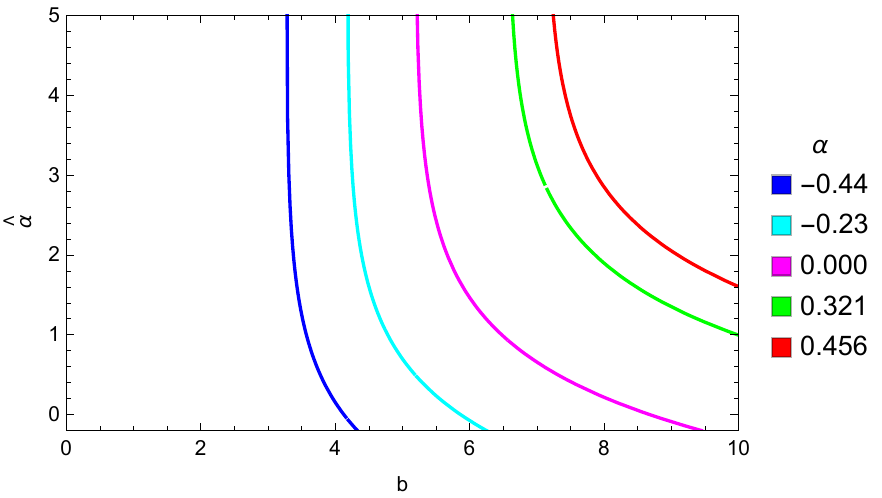}
    \caption{Strong deflection angle for the different values of dimensionless parameter $\alpha$ is plotted here. Magenta line represents the strong deflection angle for Schwarzschild black hole. }
    \label{fig:1}
\end{figure}
The term $I_R(r_m)$ is the leading term of the regular part of the integral. The mathematical expression of $I_R(r_m)$ is obtained by evaluating the integral as in \cite{tsukamoto2017deflection}. The regular integral $I_R(r_m)$ is defined as the integral with the divergent part subtracted, expressed as:
\begin{equation}
    I_R(r_m) = \int_{r_m}^{\infty} \big[f(r) - f_D(r)\big] \, dr,
\end{equation}
$f_D(r)$ is the term causing divergence in the original integral. We obtain
\begin{equation}\label{z1}
\begin{split}
I_R(r_m) &= 2 \log(6(-\sqrt{3} + 2)) + \frac{4}{9} \alpha \big(-6 + 6 \sqrt{3} + \log(3 - \sqrt{3})\big) \\
&\quad - \frac{2 \alpha^2}{81} \big(-146 + 74 \sqrt{3} + 7 \log(6) - 7 \log(3 + \sqrt{3})\big) + \mathcal{O}(M^3, \alpha^3).
\end{split}
\end{equation}
By substituting \eqref{z1} into \eqref{z}, we obtain the final expression of $\overline b$ as
\begin{equation}\label{kl}
\begin{split}
\overline{b} &=-\pi+ \log(6) + 2 \log\left(-6(-2 + \sqrt{3})\right) + \frac{1}{9} \alpha \left(-25 + 24 \sqrt{3} + \log(6) + 4 \log(3 - \sqrt{3})\right) \\
& + \frac{1}{162} \alpha^2 \left(589 - 296 \sqrt{3} - 35 \log(6) + 28 \log(3 + \sqrt{3})\right)+\mathcal{O}(M^3, \alpha^3).
\end{split}
\end{equation}
When \(\alpha=0\), the expression of \(\overline{a}\) simplifies to
\begin{equation}
\overline{a} = 1.
\end{equation} 
Similarly, the expression for \(\overline{b}\) reduces to
\begin{equation}
\overline{b} = -\pi + \log(6) + 2 \log\left(-6(-2 + \sqrt{3})\right)=-0.40023.
\end{equation}
Thus, in the limit as \(\alpha\) approaches zero, both \(\overline{a}\) and \(\overline{b}\) correspond to their respective values in the Schwarzschild case \cite{paul2020strong}.
By inserting equations \eqref{y}, \eqref{z}, and \eqref{kl} into equation \eqref{z2}, we obtain the strong deflection angle of light by the dark compact object in MOG as follows:
\begin{equation}\label{z3}
\begin{split}
\hat{\alpha} &= -\pi + \log(6) + 2 \log\left(-6(-2 + \sqrt{3})\right) \\
&\quad + \frac{1}{9} \alpha \left(-25 + 24 \sqrt{3} + 5 \log(6) + 4 \log(3 - \sqrt{3})\right) \\
&\quad + \frac{1}{162} \alpha^2 \left(589 - 296 \sqrt{3} - 35 \log(6) + 28 \log(3 + \sqrt{3})\right) \\
&\quad - \frac{1}{162} \left(162 + 18 \alpha - 7 \alpha^2 \right) \log\left(-1 + \frac{b}{b_c}\right) + \mathcal{O}(M^3, \alpha^3).
\end{split}
\end{equation}
Strong deflection angle $\hat{\alpha}$ plotted against $b$ for different values of $\alpha$ is shown in Figure \ref{fig:1}. As the value of $\alpha$ increases, the deflection angle increases. Also, the graphs of $\overline{a}$ and $\overline{b}$ are drawn for different values of $\alpha$ can be seen in Figure \ref{fig:2}. All in all, we observe that in MOG, changing values of $\alpha$ significantly impact the behavior of strong deflection angle coefficients. 

\section{Observables in strong field limit}\label{6}
 
\ 

The strong field limit approximation can be used to study the deflection angle of light involving logarithmic and constant terms. The main observable in the strong field limit is obtained by substituting the specific values of a given spacetime in the equation \cite{bozza2002gravitational} 
\begin{equation}\label{z4}
    \hat{\alpha}=-\overline{a} \hspace{0.2cm}\log\left(1- \frac{\theta D_{OL}}{b_c}\right)+ \overline{b}.
\end{equation}
The lens equation is defined in \cite{bozza2001strong} as
\begin{equation}
    \beta=\theta-\frac{D_{LS}}{D_{OS}} \Delta \alpha.
\end{equation}
where ${D}_{LS}$ denotes the distance between the lens and the source, \(\beta\) and \(\theta\) represent the angular distances from the optical axis to the source and image, respectively, and \(\Delta \hat{\alpha}_n = \hat{\alpha} - 2n \pi\) signifies the adjustment of the deflection angle after accounting for \(n\) loops of photons around the lens.

\begin{figure}[htbp] 
    \centering
    \includegraphics[width=0.49\linewidth]{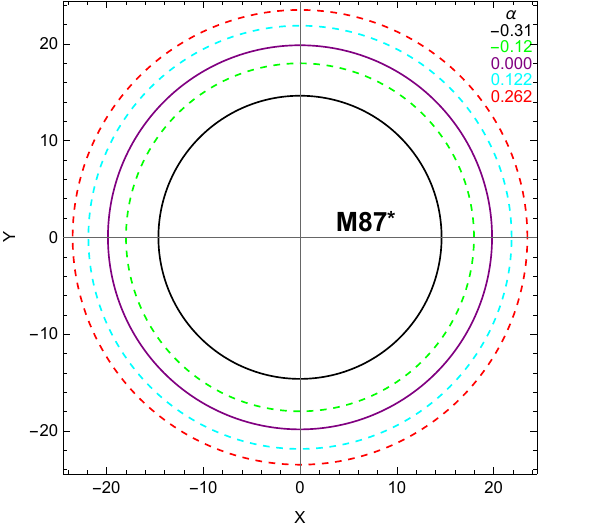}
    \includegraphics[width=0.49\linewidth]{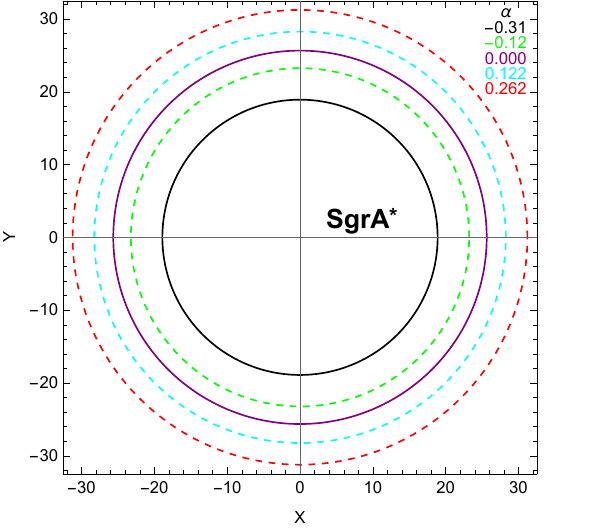}
    \caption{The Einstein rings for different values of $\alpha$ are plotted here. In the left figure, M87$^*$ acts as a lens in MOG. The right figure shows the Einstein rings when Sgr A$^*$ acts as a lens in MOG. The purple circle represents the Einstein ring for Schwarzschild spacetime.}
    \label{e1}
\end{figure}

\

To shift from the deflection angle $\hat{\alpha}$ to offset $\Delta \hat{\alpha}$, one has to calculate the $\theta^0_n$ such that $\hat{\alpha}(\theta ^0_n)=2n \pi$. By solving \eqref{z4} and using condition $\hat{\alpha}(\theta ^0_n)=2n \pi$, we obtain 
\begin{equation}\label{33}
    \theta_0^n=\frac{b_c}{D_{OL}}(1+e_n),
\end{equation}
where,
\begin{equation*}
    e_n=\exp\left(\frac{\overline b-2n \pi}{\overline a}\right).
\end{equation*}
$\Delta \hat{\alpha}_n$ can be found by expanding $\alpha(\theta)$ in power of $\theta_n^0$. By putting $\theta-\theta_n^0=\Delta \theta_n$, following expression is obtained as in \cite{bozza2002gravitational}
\begin{equation}\label{55}
\Delta \alpha_n= -\frac{\overline a D_{OL}}{b_c e_n} \Delta \theta_n
\end{equation}

\begin{table}[h]
    \centering
    \begin{tabular}{|c|c|c|c|c|c|c|c|c|}
    \hline \hline
         & \multicolumn{1}{c}{\textbf{Schwarzschild BH }} & \multicolumn{6}{|c|}{\textbf{M$87^{*}$ in Modified Gravity}} \\
         \hline  
         $\alpha$ &0 & $-0.3$ & $-0.2$ & $-0.1$ & 0.1 & 0.2& 0.3  \\ \hline
         $\theta_\infty (\mu \text{as})$ & 19.82 & 14.80 & 16.49 & 18.16 & 21.46 & 23.09 & 24.71 \\
          $s (\mu \text{as})$ & 0.0248 & 0.0075 & 0.0114 & 0.0170 &  0.0358 & 0.0510 & 0.0722 \\
          $\frac{b_{c}}{r_s}$ & 2.6 & 1.9 & 2.2 & 2.4 & 2.8 & 3.0 & 3.2  \\
          $\overline{a}$ & 1 & 0.96 & 0.98 & 0.99 & 1.01 & 1.02 & 1.03 \\
          $\overline{b}$ & -0.4002 & -1.0122  & -0.8153 & -0.6113 & -0.1821  & 0.0429 &  0.2752 \\ \hline
    \end{tabular}
\caption{This table estimates the primary observable and the strong field limit coefficients for the black hole $\text{M87}^*$ in the galaxy's center, considering various values of $\alpha$ for spacetime in modified gravity. $M = 6.5 \times 10^9 M_\odot $ and $\text{D}_{OL} = 16.8 \text{ \text{Mpc}}$ are used here. The terms $\theta_\infty$ and $s$ are defined in the text. The value $r_{ph}$ is represented as $r$ in the magnitude form. $\overline{a}$ and $\overline{b}$ are the strong field limit coefficients and $r_s$ is the Schwarzschild radius.}
    \label{Table:1}
    \end{table}

\begin{table}[h]
    \centering
    \begin{tabular}{|c|c|c|c|c|c|c|c|c|}
    \hline \hline
         & \multicolumn{1}{c}{\textbf{Schwarzschild BH }} & \multicolumn{6}{|c|}{\textbf{Sgr A$^*$ in Modified Gravity}} \\
         \hline  
         $\alpha$ &0 & $-0.3$ & $-0.2$ & $-0.1$ & 0.1 & 0.2& 0.3  \\ \hline
         $\theta_\infty (\mu \text{as})$ & 25.61 & 19.13 & 21.31 & 23.47 & 27.73 & 29.84 & 31.94 \\ 
          $s (\mu \text{as})$& 0.0319 & 0.0098 & 0.0148 & 0.0219 & 0.0462 & 0.0659 & 0.0932\\
        $\frac{b_{c}}{r_s}$ & 2.6 & 1.9 & 2.2 & 2.4 & 2.8 & 3.0 & 3.2  \\
          $\overline{a}$ & 1 & 0.96 & 0.98 & 0.99 & 1.01 & 1.02 & 1.03 \\
          $\overline{b}$ & $-0.4002$ & $-1.0122$  & $-0.8153$ & $-0.6113$ & $-0.1821$  & 0.0429 &  0.2752 \\ \hline
    \end{tabular}
    \caption{We consider Sgr A$^*$ as the lensing object, with a mass \( M = 4 \times 10^6 \, M_\odot \) and an angular diameter distance to the lens \( \text{D}_{OL} = 0.008 \, \text{Mpc} \). In this context, \( \overline{a} \) and \( \overline{b} \) represent the coefficients in the strong field limit, while \( r_s \) denotes the Schwarzschild radius.}
    \label{Table:2}
\end{table} 
\begin{figure}[ht]
    \centering
    \includegraphics[width=0.5\linewidth]{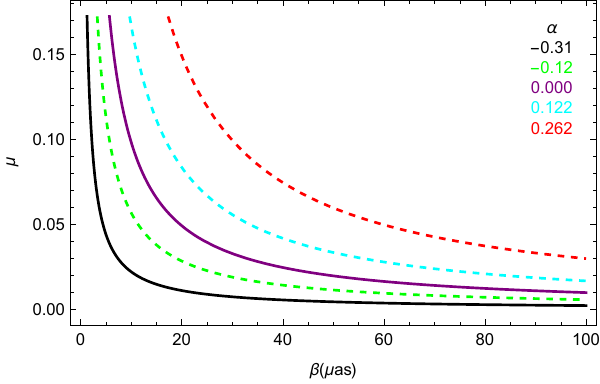}\hfill
    \includegraphics[width=0.49\linewidth]{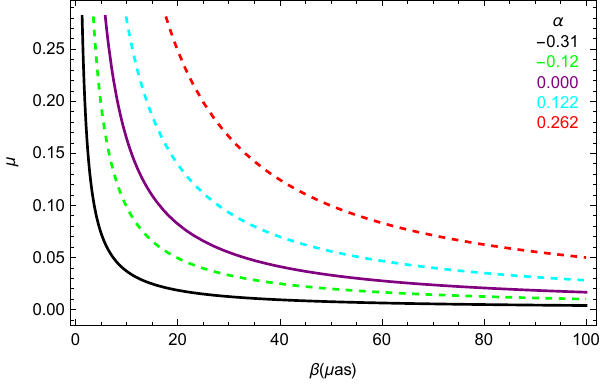}
    \caption{The magnification of the Einstein ring is plotted against various values of $\beta$ and $\alpha$. The graph illustrates two cases where black holes act as lenses in MOG: M87$^{*}$ on the left and Sgr A$^{*}$ on the right.}
    \label{m}
\end{figure}
\vspace{-5mm}
\begin{figure}[ht]
    \centering
    \includegraphics[width=0.5\linewidth]{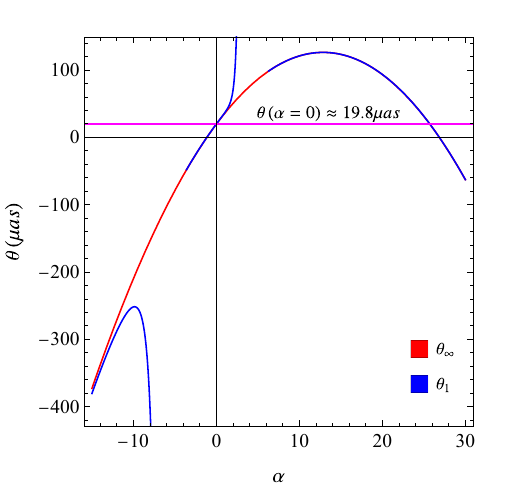}\hfill
    \includegraphics[width=0.48\linewidth]{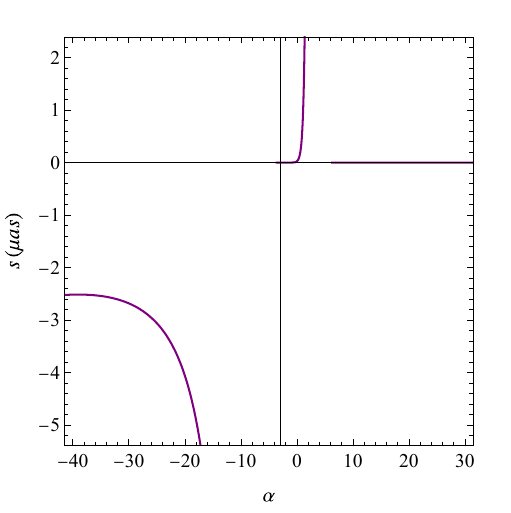}
    \caption{The behavior of the lensing coefficients: $\theta_{\infty}$ with $\theta_{1}$ (left) and $s$ (right) in MOG for different values of $\alpha$. Here, M87${^*}$ is taken as the lens. $\theta_{\infty}$ is a
    completely smooth curve. For $\theta_{1}$ and consequently $s$, there are three separate branches since each discontinuity are caused by the roots of $\overline{a}$ (\ref{y}) with occurs at $\alpha \approx -3.69$ and $\alpha \approx 6.27$.}
    \label{fig:13}
\end{figure}
Using \eqref{33} and \eqref{55} in \eqref{z4} yields 
\begin{equation}\label{66}
    \beta=(\theta_n^0+\Delta \theta_n)-\left(\frac{\overline a D_{OL}}{b_c e_n} \frac{D_{LS}}{D_{OS}}\right)  \Delta \theta_n.
\end{equation}
Using the approximation $b_c=\tan \theta D_{OL} \approx \theta D_{OL}$ along with \eqref{66}, the nth position of relativistic images will be obtained as
\begin{equation}
    \theta_n=\theta^0_n+\frac{b_c e_n (\beta-\theta^0_n)D_{OS}}{\overline a D_{LS} D_{OS}}.
\end{equation}
When $n \rightarrow \infty$ in \eqref{33}, $e_n \rightarrow 0 $ such that $b_c= \theta_\infty D_{OL}$. For $\beta=0$, when the lens, source, and object are perfectly aligned, we obtain the position of n relativistic Einstein rings as 
\begin{equation}\label{outerE}
    \theta^E_n=\frac{b_c}{D_{OL}}(1+e_n).
\end{equation}
Note that $n=1$ corresponds to the outermost relativistic ring. The magnification of the images can be calculated by using formula \cite{bozza2002gravitational},
\begin{equation}\label{22}
    \mu_n=\frac{1}{\beta}\left[\frac{b_c}{D_{OL}}(1+e_n)\left(\frac{D_{OS}}{D_{LS}} \frac{b_c e_n}{\overline a D_{OL}}\right)\right].
\end{equation}
The magnification of the Einstein ring against $\beta$ is drawn in Figure \ref{m}. It is clear from \eqref{22} and plots in Fig. (\ref{m}) that as $\beta \rightarrow0$ the magnification $\mu_n \rightarrow \infty$ which suggests that magnification is maximum in the case of a perfect alignment of lens, source, and observer. Further, \eqref{22} relates the magnification with image and source positions. 
Finally, to obtain the lensing coefficients, we treat $\theta_1$ as the outermost image and every other images are 
gathered at $\theta _ \infty$. We define two lensing coefficients as in \cite{bozza2002gravitational}
\begin{equation}
   \theta_\infty=\frac{b_c}{D_{OL}},
\end{equation}
and
\begin{equation}
    s=\theta_1-\theta_\infty.
\end{equation}
Here, $s$ is the separation between the outermost and innermost images\footnote{For MOG, the separation can be approximated by $s\approx \theta_\infty(e_n)$ only when the value of $\overline{a}$ in the exponent $e_n$ is relatively far away from zero since $e_n\rightarrow\infty$ as $\overline{a}\rightarrow0$.}, the latter forming the edge of the shadow region. Graphs of how the lensing coefficients evolve with $\alpha$ are given in Figure \ref{fig:13}.

\
We can also constrain the metric parameter $\alpha$
against past observational data to investigate how far the spherically symmetric MOG deviates from Schwarzschild spacetime. We will conduct it mainly using the measured radius of Einstein rings along with the estimated total mass enclosed within the ring. For a start, we assume that the observed Einstein radius equals the outermost image value, (\ref{33}). Substituting (\ref{bc}) into (\ref{33}) and solving for $M$ gives
\begin{equation}\label{M1}
M_1=\dfrac{\theta_E D_{OL}}{
  \left(1+e_n\right)
  \left(3\sqrt{3}+\frac{5}{2}\sqrt{3}\alpha-\frac{7}{24\sqrt{3}}\alpha^2\right)}
  +\mathcal{O}(M^3,\alpha^3)
\end{equation} 
Since a single $\overline{a}$ and $\overline{b}$ value yields two possible $\alpha$ values and accounting for the possibility of either a primary or secondary image forming, we can obtain up to four range of constrains for each compact body. The following constraints are estimated with (\ref{outerE}) and (\ref{M1}) and up to 3 decimal places.  

\begin{figure}[ht]
    \centering
    \includegraphics[width=0.5\linewidth]{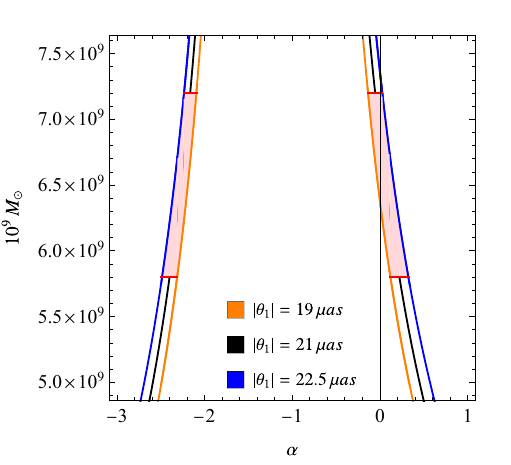}\hfill
    \includegraphics[width=0.5\linewidth]{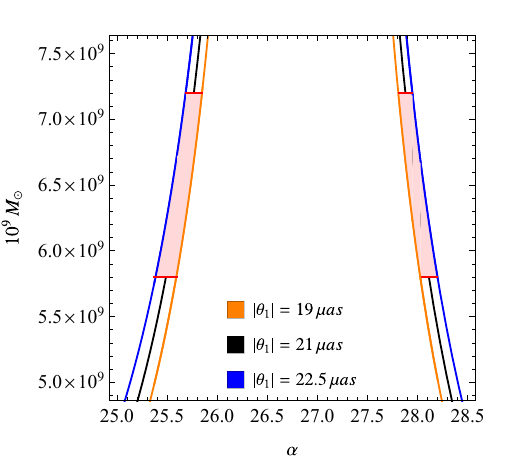}
    \caption{Constraints of $\alpha$ taking M87 as a lens. In the left figure, the right branches represent positive $\theta_{1}$ values whilst the left branches are the negative ones. It is the reversed way in the right figure. The shaded regions contains the values of the estimated mass up to $1-\sigma$ bound.}
    \label{M87Con}
\end{figure}

Taking M87${^*}$ as the lensing body and using observational results of \cite{event2019first}, we have the following constraints up to the $1-\sigma$ level,
\begin{subequations}\label{M87range}
\begin{align}
&\theta_E^{+}\,(\alpha<\alpha_{crit}): -0.142 \lesssim \alpha \lesssim 0.338, \\
&\theta_E^{+}\,(\alpha>\alpha_{crit}): 25.371 \lesssim \alpha \lesssim 25.855, \\
&\theta_E^{-}\,(\alpha<-1): -2.496 \lesssim \alpha \lesssim -2.089, \\
&\theta_E^{-}\,(\alpha>\alpha_{crit}): 27.803 \lesssim \alpha \lesssim 28.210, 
\end{align}
\end{subequations}
and for Sgr A${^*}$, up to $68\%$ CI \cite{akiyama2022first},
\begin{subequations}\label{SgrArange}
\begin{align}
&\theta_E^{+}\,(\alpha<\alpha_{crit}): -0.296 \lesssim \alpha \lesssim 0.312, \\
&\theta_E^{+}\,(\alpha>\alpha_{crit}): 25.397 \lesssim \alpha \lesssim 26.010, \\
&\theta_E^{-}\,(\alpha<-1): -2.475 \lesssim \alpha \lesssim -1.953, \\
&\theta_E^{-}\,(\alpha>\alpha_{crit}): 27.667 \lesssim \alpha \lesssim 28.189, 
\end{align}
\end{subequations}
Positive valued Einstein radius $\theta_{E}^+$ represents a primary image whilst the negative valued $\theta_{E}^-$ describes a secondary image with opposite parity.
From (\ref{M87range}) and (\ref{SgrArange}), we see that if the observed Einstein ring is taken to be a primary image, it indicates that the lensing body is either a black hole or a regular dark compact object. Additionally, only horizonless compact objects including a naked singularity are capable of producing a secondary image. Constraints for M87${^*}$ are illustrated graphically in Figure \ref{M87Con}.

\
Next, we do the constraint test on other Einstein rings lensed by galaxies and aim to probe how far could MOG describe a galaxy lens assumed as a circularly symmetric point mass in flat cosmology. Strong lensing are frequent 
at cosmological distances \cite{saha}. The total estimated mass enclosed by a given Einstein ring in MOG can be determined by simply rearranging (\ref{mogring}),
\begin{equation}\label{M2}
M_2=\dfrac{\theta^2_{E}\mathcal{D}}{4(1+\alpha)}    
\end{equation}
where $\theta_E$ is the Einstein radius and 
$\mathcal{D}$ is the effective angular diamter distance given by the ratio $\frac{D_{OL}D_{OS}}{D_{LS}}$. There is only one effective range of constraint for each Einstein ring since (\ref{M2}) has a linear $\alpha$ term and the squared $\theta_{E}$ term yields the same magnitude regardless if the image is primary ($\theta_{E}>0$) or secondary ($\theta_{E}<0$). 

Astronomical data typically measures the distance between celestial objects and Earth-bound observers in terms of redshift $z$ to account for the expanding Universe. In our case, we need to convert the measured redshift of the lens and source into their respective angular diameter distances. The angular diameter distances between the observer and either the lens or the source $D_{OL}$ and $D_{OS}$ can be determined by \cite{saha},
\begin{equation}\label{distLSz1}
\dfrac{D_H}{(1+z)}\int_{0}^{z} \dfrac{dz'}{E(z')},
\end{equation}
where the function $E(z)=\sqrt{\Omega_M (1+z)^3+\Omega_k (1+z)^3+ \Omega_{\Lambda}}$ contains the density parameters 
for matter, curvature and ``dark energy" in order, and
obey the relation $\Omega_M+\Omega_k+\Omega_\Lambda=1$. 
$D_H$ is the Hubble distance defined by $D_H=\frac{c}{H_0}$,
which we will take it to be approximately $2998 h^{-1}$ Mpc.
Note that $h$ is a numerical value associated with the Hubble constant, $H_0$ \cite{cosmoh}. 

The angular diameter distance between the lens and the source
in flat cosmology $(\Omega_k=0)$ is given by,
\begin{equation}\label{distLSz2}
D_{LS}=\dfrac{D_H}{1+z_{S}}\left[ \int_{0}^{z_{S}} \dfrac{dz'}{E(z')} - \int_{0}^{z_{L}} \dfrac{dz'}{E(z')} \right],
\end{equation}
with $z_S$ and $z_L$ being the redshift of the source and the lens respectively. In general, $D_{LS} \neq D_{OS} - D_{OL} $.
We summarized the constraints of several Einstein rings modeled with the aforementioned assumption in Table \hyperref[Table:3]{3}. Constraints are estimated via (\ref{M2}), (\ref{distLSz1}) and (\ref{distLSz2}).

\begin{table}[ht]
\scalebox{0.9}{\begin{tabular}{l c c c c c c}
\hline
Einstein Ring & $z_L$ & $z_S$ & $\mathcal{D}$ ($h^{-1}$ Mpc)
& $\theta_{E}$ ($''$) & $M\,(10^{12} h^{-1} M_{\odot}$) & Constraints on $\alpha$ \\
\hline
8 O'Clock Arc \cite{8oClockArc} & 0.38 & 2.73 & 993.67 & $(3.32 \pm 0.16)$ & $(1.35 \pm 0.12)$ & $(-0.179,0.216)$ \\
Clone \cite{Lin_2009} & 0.422 & 2.00 & 1168.14 & $(3.82 \pm 0.03)$ & $(2.10 \pm 0.03)$ & $(-0.034,0.029)$ \\
Elliot Arc \cite{Buckley} & 0.3838 & 0.9057 & 1475.16 & $(7.53 \pm 0.25)$ & $(10.3 \pm 0.7)$ & $(-0.130,0.101)$ \\
Canarias \cite{bettinelli2016canarias} & 0.581 & 1.165 & 2277.31 & $(2.16 \pm 0.13)$ & 
$(1.30 \pm 0.16)$ & $(-0.213,0.292)$ \\
\hline
\end{tabular}}
\caption{Summary of estimates for bounds of the parameter $\alpha$ in spherically-symmetric MOG from several observed Einstein rings. The symbol $''$ stands for arcseconds. We assume $\Omega_M=0.3,\,\Omega_{\Lambda}=0.7$ and $H_0=100\,h$ km s$^{-1}$ Mpc$^{-1}$ for all rings.}
\end{table}\label{Table:3}

\subsection{Computations and Results}

\ 

We modeled M87$^{*}$ as a lens within the framework of MOG. The lensing coefficients $\theta_{\infty}$ and $s$, along with other parameters such as $\frac{b_c}{r_s}$ (where $b_c$ is the critical impact parameter of the light ray and $r_s$ is the Schwarzschild radius), $\overline{a}$, and $\overline{b}$, were computed. These values are presented in Table \ref{Table:1}. The same analysis was performed for Sgr A$^{*}$, treating it as a lensing body in MOG, with the results provided in Table \ref{Table:2}. Additionally, graphs of $\theta_{\infty}$, $\theta_{1}$, and $s$ were plotted for various values of $\alpha$, as shown in Figure \ref{fig:13}.  The $\theta_n$ values behave differently from $\theta_{\infty}$. Since ($\ref{y}$) is quadratic, the two zeroes of $\overline{a}$ cause the $e_{n}$ term in (\ref{33}) to diverge, giving two jump discontinuities. Einstein rings are formed when the lens, source, and observer are aligned perfectly. The outermost Einstein rings for the varying $\alpha$ are plotted here using \eqref{33} in Figure \ref{e1}. Both M87${^*}$ and Sgr A${^*}$ were taken as the lensing body.

\
On the aspect of how the lensing coefficients changes with $\alpha$ (see Figure \ref{fig:13}), the positive-value $\theta_{\infty}$ representing the primary image will reach a maximum value at a certain $\alpha$. For M87${^*}$, this is about $126\,\mu as$ at $\alpha=12.857$. Moving away from this $\alpha$ value in both directions, the radius of the Einstein ring of the primary image decreases rapidly and then continue in this manner for negative $\theta_{\infty}$ values. Negative $\theta_{\infty}$ represents the radius of the inverted secondary image. The separation between the outermost and innermost image is negligible at the rightmost branch, that is $\alpha$-values to the right of the larger root ($\alpha=6.27$) and close to $\alpha=0$ (middle branch). From the middle branch, $\theta_{1}$ diverges to either positive or negative infinity when approaching the larger or smaller $\alpha$ value corresponding to each $\overline{a}$ root respectively. For the leftmost branch which contains negative $\alpha$-values, we observe that the separation $s$ between the secondary images are typically larger than the primary image counterpart. 

\
We find the constraints on metric parameter $\alpha$ are stronger for M87${^*}$ in all four possible ranges. These constraints all lie within parts of the three branches of $\theta_{1}$ with separation from $\theta_{\infty}$ much less than unity implying a relatively good fit. 
From Table \hyperref[Table:3]{3}, the Clone Einstein ring gives the tightest constraint and the Canarias ring shows the largest deviation from $\alpha=0$. Simply put, a smaller error on the calculated mass appear to give more restriction on $\alpha$.

\begin{table}
\begin{tabular}{l c c c c c c}
\hline
Galaxy & \( M(M_\odot) \) & \( D_{OL} \) (Mpc) & \( \frac{M}{D_{OL}} \) & & Schwarzschild $\Delta T ^{s}_{2,1}$ &  \\
 &  &  &  & $\alpha=-0.1$ & $\alpha=0$ &$\alpha=0.1$  \\
\hline
Milky Way     & \( 4.0 \times 10^6 \) & 0.008  & \( 2.471 \times 10^{-11} \) &9.804 & 10.699 & 11.587\\
M87           & \( 6.5 \times 10^9 \) & 16.8   & \( 1.758 \times 10^{-11} \) &15931.23 & 17385.67 & 18828.84\\
NGC 4472      & \( 2.54 \times 10^9 \) & 16.72  & \( 7.246 \times 10^{-12} \)& 6225.34& 6793.78 & 7357.73\\
NGC 1332      & \( 1.47 \times 10^9 \) & 22.66  & \( 3.094 \times 10^{-12} \)& 3602.908 & 3931.84 & 4258.11\\
NGC 4374      & \( 9.25 \times 10^8 \) & 18.51  & \( 2.383 \times 10^{-12} \)& 2267.14 & 2474.11 & 2679.49\\
NGC 1399      & \( 8.81 \times 10^8 \) & 20.85  & \( 2.015 \times 10^{-12} \)& 2159.294 & 2356.43 & 2552.03 \\
NGC 3379      & \( 4.16 \times 10^8 \) & 10.70  & \( 1.854 \times 10^{-12} \)&1019.60 & 1112.68 & 1205.04 \\
NGC 4486B     & \( 6 \times 10^8 \)    & 16.26  & \( 1.760 \times 10^{-12} \)& 1470.58 & 1604.83 & 1738.05\\
NGC 1374      & \( 5.90 \times 10^8 \) & 19.57  & \( 1.438 \times 10^{-12} \)& 1446.07& 1709.08 & 1578.08 \\
NGC 4649      & \( 4.72 \times 10^9 \) & 16.46  & \( 1.367 \times 10^{-12} \)& 11568.52 & 12624.671 & 13672.64\\
NGC 3608      & \( 4.65 \times 10^8 \) & 22.75  & \( 9.750 \times 10^{-13} \)& 1139.70 & 1243.74 & 1346.99 \\
NGC 3377      & \( 1.78 \times 10^8 \) & 10.99  & \( 7.726 \times 10^{-13} \)& 436.27& 476.099 & 515.62 \\
NGC 4697      & \( 2.02 \times 10^8 \) & 12.54  & \( 7.684 \times 10^{-13} \)&495.09 & 540.29 & 585.14 \\
NGC 5128      & \( 5.69 \times 10^7 \) & 3.62   & \( 7.498 \times 10^{-13} \)& 139.46 & 152.19 & 168.82 \\
NGC 1316      & \( 1.69 \times 10^8 \) & 20.95  & \( 3.848 \times 10^{-13} \)& 414.21 & 452.03 & 489.55 \\
NGC 3607      & \( 1.37 \times 10^8 \) & 22.65  & \( 2.885 \times 10^{-13} \)& 335.78 & 366.44 & 396.85 \\
NGC 4473      & \( 0.90 \times 10^8 \) & 15.25  & \( 2.815 \times 10^{-13} \)& 220.59 &240.72 & 260.71 \\
NGC 4459      & \( 6.96 \times 10^7 \) & 16.01  & \( 2.073 \times 10^{-13} \)& 170.59& 186.16 & 201.61\\
M32           & \( 2.45 \times 10^6 \) & 0.8057 & \( 1.450 \times 10^{-13} \)& 6.01&6.55 & 7.10 \\
NGC 4486A     & \( 1.44 \times 10^7 \) & 18.36  & \( 3.741 \times 10^{-14} \)& 35.29 &38.52 & 41.70 \\
NGC 4382      & \( 1.30 \times 10^7 \) & 17.88  & \( 3.468 \times 10^{-14} \)& 31.86 & 34.77 & 37.66 \\
\hline
\end{tabular}
\caption{Estimation of time delay between the first and second relativistic image for supermassive black holes at the center of nearby galaxies. Mass (\(M\)) and distance (\(D_{OL}\)) are given in units of solar mass and Mpc, respectively  \cite{kormendy2013coevolution}. Time delays are measured in minutes.}
\label{Table:4}
\end{table}

\section{\textbf{Time delay in SDL}}
Using the approach outlined in \cite{bozza2004time}, the time delay between the relativistic images is calculated. This delay occurs because photons follow distinct winding paths around the black hole, leading to a time difference between images formed on opposite sides of the lens. Given that the images are highly de-magnified and separated by only a few microarcseconds ($\mu$as), it is crucial to distinguish the outermost image from the others. For spherically symmetric black holes, when both images are on the same side of the source, the time delay between the first and second relativistic images is given by \cite{bozza2004time} as:

\begin{equation}
\Delta T^s_{2,1} = 2 \pi b_c = 2 \pi D_{OL} \theta_{\infty}.
\end{equation}
The time delay for twenty-one different galaxies is calculated in Table \ref{Table:4}. For  Sgr A$^{*}$, the time delay will reach $\sim$9.804 min and $\sim$11.587 min for $\alpha=-0.1$ and $\alpha=0.1$ in modified gravity, respectively. It can be observed that the positive values of $\alpha$ cause photons to experience greater gravitational influence, thereby increasing the delay between different relativistic images in MOG while the reverse for negative values of $\alpha$. Time delay for M87$^{*}$ is not greater in MOG but for other galaxies, it is significant and can studied easily.
\section{Conclusion}\label{7}

\ 
The study of modified gravity theories has gained much attention in the past few years. The STVG theory has been an interesting topic to analyze after its proposal by Moffat. Later, the shadow patterns, orbit formation, and accretion disks are studied in Modified Gravity.  In this paper, we have used the line element of spacetime in modified gravity and calculated the weak deflection of photons originating from the asymptotic region. The observables in the weak field limit are discussed here by using M87$^*$ as a lens with $M = 6.5 \times 10^9 M_\odot $ and $\text{D}_{OL} = 16.8 \text{ \text{Mpc}}$ which developed a deep understanding of magnification of relativistic images in weak field limit. A distortion parameter is also plotted to analyze the distortion of images in weak deflection limits and their signed sum is zero obeying Virbhadra's hypothesis \cite{virbhadra2022distortions}. 

\ 
Moreover, the bending angle in strong gravity is computed, and observables are discussed. We find that larger $\alpha$
produces larger magnification values at large $\beta$ for each compact object but has a smaller magnification increasing rate in the $\beta\rightarrow0$ direction.
Due to our second-order approximation, the separation between the outermost and image diverges near the $\alpha$ values corresponding to each roots of the lensing coefficient $\overline{a}$. Separation for the secondary images are larger than the primary ones. Constraints test on $\alpha$ are conducted with observational data of the supermassive black holes M87$^*$ and Sgr A$^*$ and four different Einstein rings produced by a system of two redshifted galaxies. We assume a simple circularly symmetric 
point mass model for the galactic systems and found that the deviation of $\alpha$ increases with the error from calculating the mass. Several range of constraints are possible using the mass equation derived from strong field observables. So this suggests that modified gravity accommodates black holes and other compact objects, with unique observational features that distinguish it from classical GR black hole predictions. This analysis provides a method to identify horizonless objects through specific observational constraints. 

In addition, we analyzed the time delays for twenty-one different galaxies, as presented in Table \ref{Table:3}. The results indicate that time delays increase when the free parameter $\alpha$ is positive and decreases when $\alpha$ is negative, hence providing a better way to understand gravitational lensing in the strong deflection limit of the dark compact objects in MOG.  In conclusion, dark compact objects in modified gravity provide a profound insight into strong gravitational lensing.
\bibliography{ref}
\bibliographystyle{IEEEtran}
\end{document}